\documentclass[aps,prb,twocolumn,groupedaddress,showpacs]{revtex4-1}
\usepackage{graphicx}

\begin{document}

\title{Evidence for electron-phonon interaction in Fe$_{1-x}$M$_{x}$Sb$%
_{2}$ (M=Co, Cr) single crystals}
\author{N. Lazarevi\'c, Z. V. Popovi\'c}
\affiliation{Center for Solid State Physics and New Materials, Institute of Physics,
Pregrevica 118, 11080 Belgrade, Serbia}
\author{$^{\ast }$Rongwei Hu, C. Petrovic}
\affiliation{Condensed Matter Physics and Materials Science Department, Brookhaven
National Laboratory, Upton, New York 11973-5000, USA}

\begin{abstract}
We have measured polarized Raman scattering spectra of the
Fe$_{1-x}$Co$_{x}$Sb$_{2}$ and Fe$_{1-x}$Cr$_{x}$Sb$_{2}$ (0$\leq x\leq $0.5)
single crystals in the temperature range between 15 K and 300 K. The highest energy $B_{1g}$ symmetry mode shows significant line asymmetry due to phonon mode coupling width electronic background. The coupling constant achieves the highest value at about 40 K and after that it remains temperature independent. Origin of additional mode broadening is pure anharmonic. Below 40 K the coupling is drastically reduced, in agreement with transport properties measurements. Alloying of FeSb$_2$ with Co and Cr produces the B$_{1g}$ mode narrowing, i.e. weakening of the electron-phonon interaction. In the case of A$_{g}$ symmetry modes we have found a significant mode mixing.
\end{abstract}

\pacs{ 78.30.Hv; 72.20.-; 75.20.-g; 73.63.-b; }
\maketitle

\section{Introduction}

FeSb$_{2}$ is a narrow-gap semiconductor which attracted a lot of attention
because of its unusual magnetic,\cite{1} thermoelectric\cite{2} and
transport properties.\cite{3} The magnetic susceptibility of FeSb$_{2}$ is
nearly constant at low temperatures with paramagnetic to diamagnetic
crossover at around 100 K for a field applied along the $c$-axis, similar to
FeSi.\cite{4} The electrical resistivity along the $a$- and $b$ - axes shows
semiconducting behavior with rapid increase for T $<$ 100 K. Along the $c$
-axis resistivity exhibits a metal to semiconductor transition at around 40
K.\cite{1,4} Based on the measurements of the
electrical resistivity, magnetic susceptibility, thermal expansion, heat
capacity and optical conductivity the FeSb$_{2}$ has been characterized as a
strongly correlated semiconductor.\cite{1,2,3,4,5,6} It was also shown that
FeSb$_{2}$ has colossal Seebeck coefficient $S$ at 10 K and the largest
power factor $S^{2}\sigma $ ever reported.\cite{2} Thermal
conductivity $\kappa $ of FeSb$_{2}$ is relatively high
and is dominated by phonons around 10 K with phonon mean free path $l_{ph}\sim 10^{2}\mu$m several orders of magnitude larger than electronic mean free path.\cite{2}

In the recent room temperature study we have observed, for the first time, all six Raman active modes of
FeSb$_{2}$ predicted by theory.\cite{7} Racu \emph{et al}.\cite{8} measured polarized Raman scattering spectra of FeSb$
_{2}$ single crystals below room temperature and found only anharmonicity of
A$_{g}$ and B$_{1g}$ symmetry modes with no additional electron-phonon coupling.

In this work we have measured at different temperatures polarized Raman scattering
spectra of pure FeSb$_{2}$ single crystals and FeSb$_{2}$ crystals alloyed with Co and Cr. The B$_{1g}$ mode asymmetry and broadening is analyzed using Breit-Wigner-Fano profile model. The coupling between single phonon and the electronic background is drastically reduced for temperatures bellow 40 K, fully in agreement with transport properties measurements.\cite{3} Alloying of FeSb$_{2}$ with Co and Cr also reduces the coupling, i.e. leads to the B$_{1g}$ mode narrowing. We have also observed strong A$_{g}$ symmetry mode mixing.
\section{Experiment}

Single crystals of FeSb$_{2}$, Fe$_{1-x}$Co$_{x}$Sb$_{2}$ and Fe$_{1-x}$Cr$_{x}$Sb$_{2}$ (0$<x\leq$0.5) were grown using the high-temperature flux method, which is described in details in Refs.\cite{9,10} Sample
structure and composition were determined by analyzing the powder X-ray
diffraction data of Fe(Co,Cr)Sb$_{2}$ single crystals collected
using a Rigaku Miniflex diffractometer with Cu K$_{\alpha }$ radiation.\cite{4}
The samples stoichiometry was determined by an
energy dispersive JEOL JSM-6500 SEM microprobe. Analysis of several
nominal x=0.25 samples showed that the uncertainty in Co and Cr
concentrations among samples grown from different batches was
$\Delta$x=0.04. The Raman scattering measurements were performed using
Jobin Yvon T64000 Raman system in micro-Raman configuration. The 514.5 nm
line of an Ar$^{+}$/Kr$^{+}$ mixed gas laser was used as an excitation source. Focusing of
the laser beam was realized with a long distance microscope objective
(magnification $50\times $).  We have found that laser power level of 0.02 mW on
the sample is sufficient to obtain Raman signal and, except signal to noise
ratio, no changes of the spectra were observed as a consequence of laser
heating by further lowering laser power. The corresponding excitation power
density was less then 0.1 kW/cm$^{2}$. All Raman scattering measurements presented in this work were performed using the (10$\bar{1}$) plane of FeSb$_{2}$ orthorhombic crystal structure.
Low temperature measurements were performed between 15 K and 300 K using KONTI CryoVac continuous Helium flow cryostat with 0.5 mm thick window.

\section{Results and discussion}

FeSb$_{2}$ crystallizes in the orthorhombic marcasite-type structure of the
centrosymetric Pnnm (D$_{2h}^{12}$) space group, with two formula units (Z=2)
per unit cell.\cite{4} Basic structural unit is built up of Fe ion surrounded
by deformed Sb octahedra. These structural units form edge sharing chains
along the $c$- axis. According to the factor-group analysis there are 6 Raman active
modes (2A$_{g}$+2B$_{1g}$+B$_{2g}$+B$_{3g}$), which were observed and
assigned in our previous work.\cite{7} The A$_{g}$
and B$_{1g}$ symmetry modes are bond stretching vibrations, whereas the B$_{2g}$ and B$_{3g}$ symmetry modes represent librational ones.\cite{12}

 \begin{figure}
\includegraphics[width=0.3\textwidth]{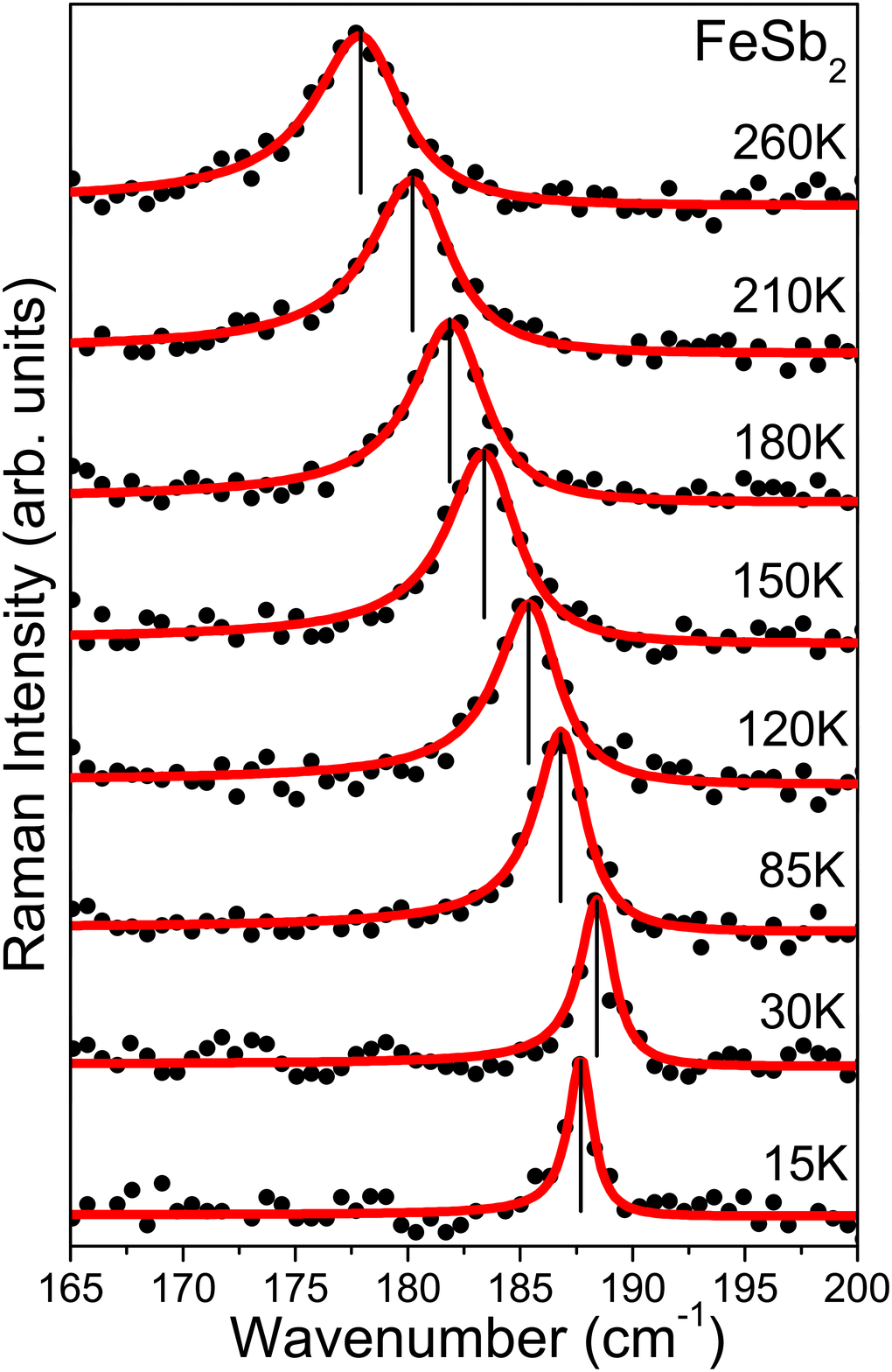}
\caption{The Raman scattering spectra of FeSb$_2$ single crystals in the (x$^,$y) configuration (B$_{1g}$ symmetry modes) measured at different temperatures.}
\label{fig2}
\end{figure}

\begin{figure}
\includegraphics[width=0.43\textwidth]{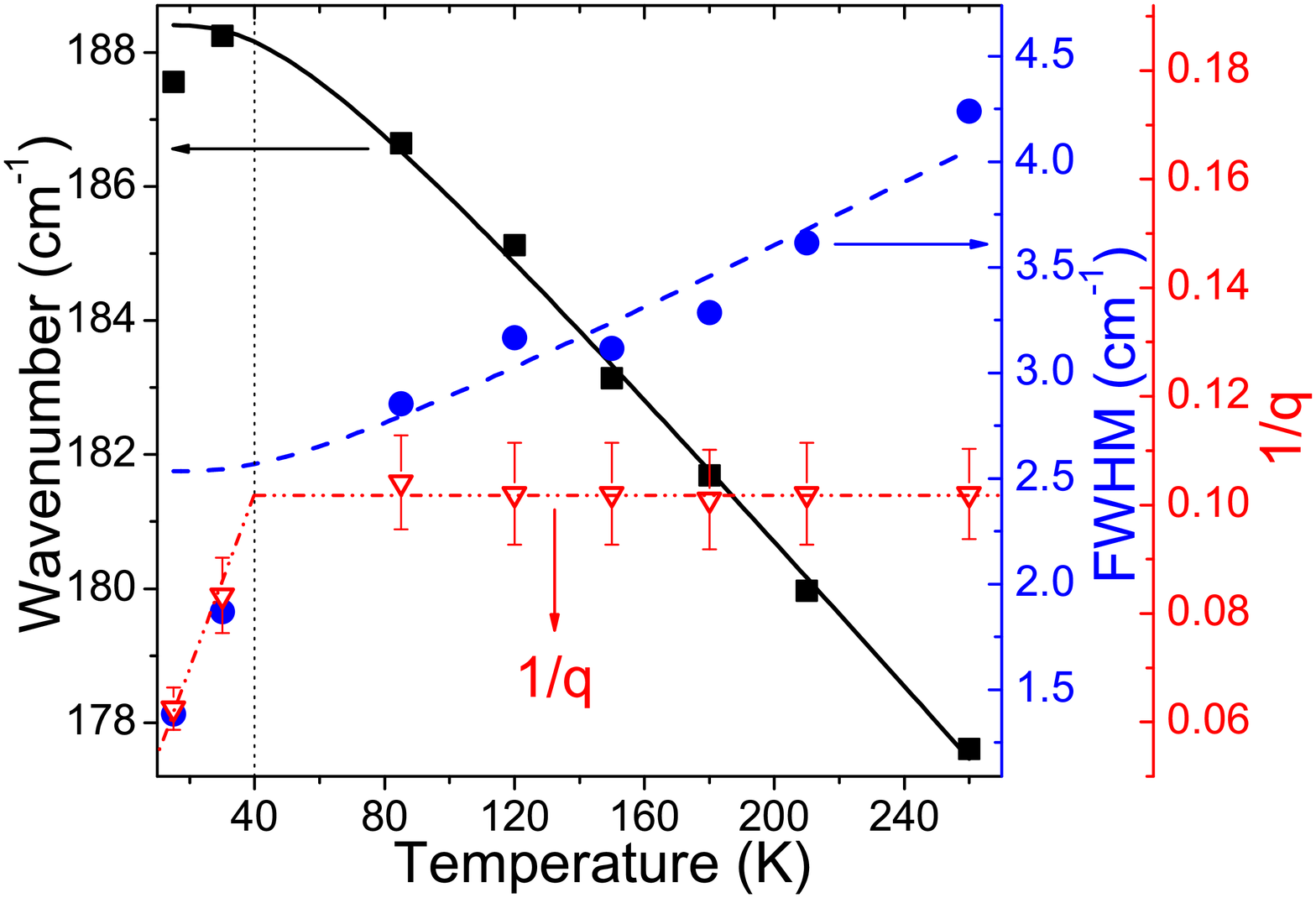}
\caption{Wavenumber, FWHM and the degree of coupling (1/$q$) of the B$_{1g}$ mode as a function of temperature for FeSb$_2$ sample.}
\label{fig3}
\end{figure}

 Fig. \ref{fig2} shows Raman scattering spectra of FeSb$_{2}$ single crystals in the (x$^{,}$y) configuration ($x^{,}=\frac{1}{\sqrt{2}}[101]$, $y=[010]$)\cite{7} measured at different temperatures in the spectral range of the highest energy B$_{1g}$ symmetry mode. For this configuration B$_{1g}$ and B$_{3g}$ modes are Raman active, see Ref \onlinecite{7}. One can notice an asymmetry of the B$_{1g}$ mode towards lower wave numbers. This broad, asymmetric structure is analyzed using a Breit-Wigner-Fano (BWF) interference model.\cite{19,20} The resonance usually involves an interference between Raman scattering from continuum excitations and that from a discrete phonon, provided two Raman-active excitations are coupled. The BWF model line shape is given by:
\begin{equation}
I=I_{0}\frac{(1-\epsilon/q)^{2}}{1+\epsilon^{2}},
\label{6}
\end{equation}
where $\epsilon=(\omega -\omega _{p})/(\Gamma/2)$ and $1/q$ is the degree of coupling which describes the departure of the lineshape from a symmetric Lorentzian function. The $I_{0}$ is the intensity and $\omega _{p}$ and $\Gamma /2$ are the real and imaginary part of phonon self energy, respectively. The spectra calculated using Eq. ($\ref{6}$) are shown as solid lines in Fig. \ref{fig2}.
\begin{table}
\caption{Parameters obtained by fitting of the B$_{1g}$ symmetry mode  spectra of pure FeSb$_2$ with the BWF line shape model.}
\label{tab.1}
\begin{ruledtabular}
\begin{tabular}{cccc}
Temperature (K) & $\omega_p$ (cm$^{-1}$) & $\Gamma$ (cm$^{-1}$) & $q$ \\
\hline
15 & 187.7(2) & 1.4(3) & 16(1) \\
30 & 188.4(1) & 1.8(3) & 12(1) \\
85 & 187.0(1) & 2.7(2) & 9.6(8) \\
120 & 185.6(1) & 3.3(3) & 9.8(9) \\
150 & 183.6(1) & 3.8(3) & 9.8(9) \\
180 & 182.1(1) & 3.9(3) & 9.9(9) \\
210 & 180.4(1) & 4.3(3) & 9.8(9) \\
260 & 178.1(1) & 4.7(3) & 9.8(8) \\
\end{tabular}
\end{ruledtabular}
\end{table}
The best fit parameters are presented in Table \ref{tab.1}. Decrease of $q$  indicates an increase in electron-phonon coupling. Nearly the same value for $q$ above 40 K (Table \ref{tab.1}) corresponds to temperature independent electron-phonon interaction contribution (see Fig. \ref{fig3}). These results are completely in agreement with the transport properties measurements which also showed that carriers concentration rapidly decreases for T$<$40 K, and is nearly constant above this temperature.\cite{3}

\begin{figure}
\includegraphics[width=0.45\textwidth]{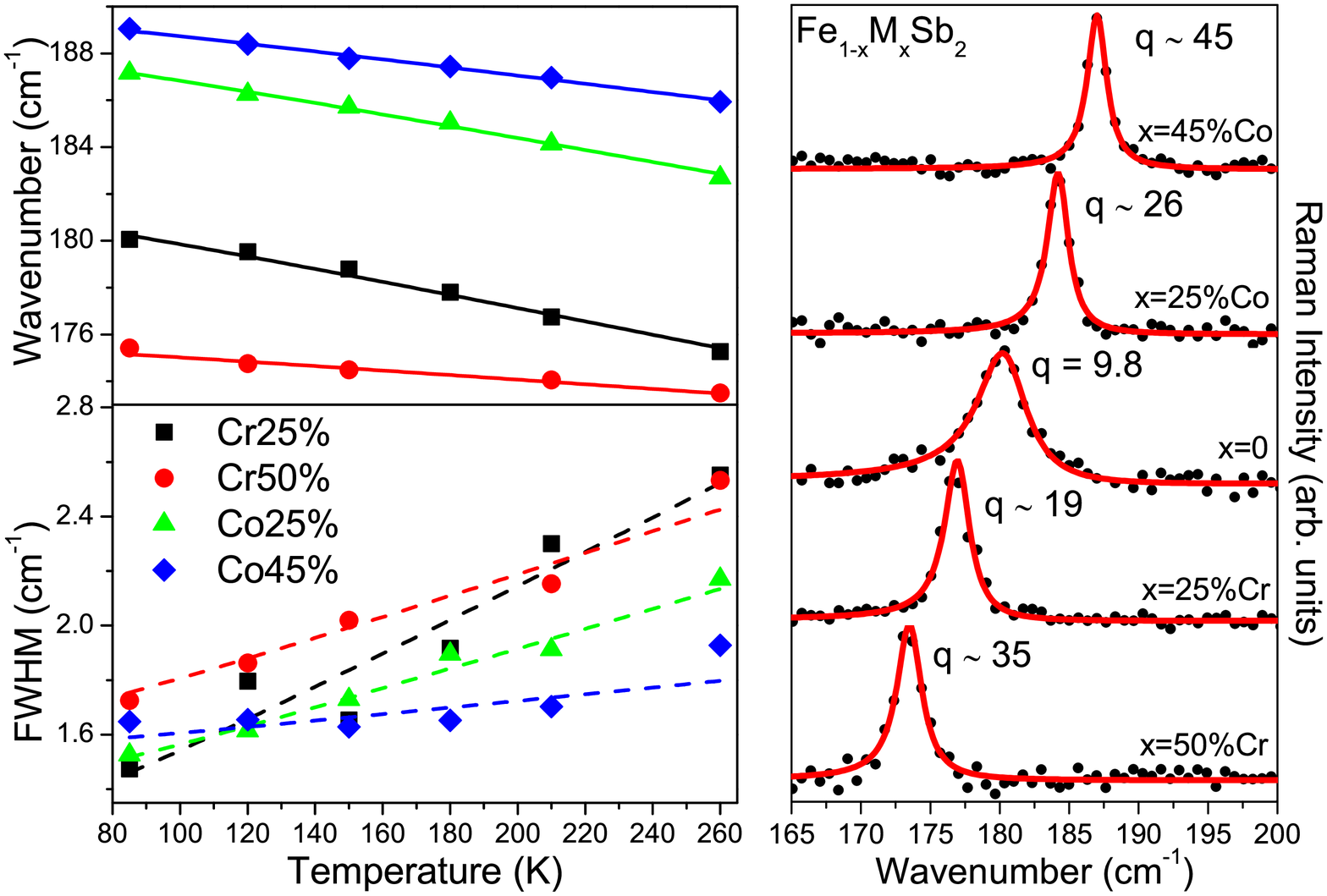}
\caption{Wavenumber and FWHM as a function of temperature (left panel) and the BWF analysis at 210 K (right panel) of the B$_{1g}$ mode for Fe$_{1-x}$(Co,Cr)$_x$Sb$_2$ alloy samples.}
\label{fig4}
\end{figure}

In general, structural disorder, isotopic and/or anharmonic effects and electron-phonon interaction can cause a change of linewidth, among which only the last two can introduce temperature dependence. Fig. \ref{fig3} shows temperature dependance of energy and linewidth of the B$_{1g}$ mode for pure FeSb$_2$ sample experimentally obtained as peak position and full width at half maximum (FWHM) of the Raman mode, respectively. As can be seen from Fig. \ref{fig3}, in the temperature range between 15 K and 40 K, the linewidth drastically increases with temperature increase. Because the phonon linewidth change due to the phonon-phonon interactions (anharmonicity) is usually very small at low temperatures we concluded that dramatic change of B$_{1g}$ mode linewidth of FeSb$_2$ below 40 K comes from strong temperature dependent electron-phonon interaction. Support for this conclusion we found in a perfect mapping of the FWHM and $1/q$ temperature dependance for T$<$40 K and the transport properties measurements,\cite{3} which shows dramatic carrier concentration decrease for T$<$40 K. At higher temperatures (in our case above 40 K) major contribution to the temperature dependance of the linewidth comes from the phonon-phonon interaction because the electron-phonon contribution for T$>$40 K is temperature independent ($1/q\sim const.$).

Having this in mind, for T$>$40 K we consider energy and linewidth change of the B$_{1g}$ mode vs. temperature as pure temperature induced anharmonic effect. Influence of the anharmonic effects on the Raman mode linewidth and energy can be taken into account via three-phonon processes:\cite{17,18}
\begin{equation}
\Gamma(T)=\Gamma_0+A\bigg(1+\frac{2}{e^{x}-1}\bigg),
\label{10}
\end{equation}
where $\Gamma_0$ includes intrinsic linewidth, structural disorder, isotopic effect and temperature independent electron-phonon interaction contribution. A is the anharmonic
constant and $x=\hbar\Omega_0/2k_{B}T$.

Phonon energy is given by:
\begin{equation}
\Omega (T)=\Omega_0-C\bigg(1+\frac{2}{e^{x}-1}\bigg),
\label{4}
\end{equation}
where $\Omega_0$ is temperature independent contributions, C is the anharmonic constant.\cite{17,18}
Eq. (\ref{10}) and Eq. (\ref{4}) give a rather good fit (dashed and solid lines in the Fig. \ref{fig3}, respectively) of the experimental data  for the temperature region above 40 K. Fit parameters are presented in the Table \ref{tab.2}.

\begin{table}
\caption{Best fiting parameters for FWHM and wavenumber temperature dependance of the B$_{1g}$ symmetry mode using Eq. (\ref{10}) and (\ref{4}).}
\label{tab.2}
\begin{ruledtabular}
\begin{tabular}{ccccc}
Compound & $\Gamma_0$ (cm$^{-1}$) & A (cm$^{-1}$) & $ \Omega_0$ (cm$^{-1}$) & C (cm$^{-1}$) \\ \hline
FeSb$_2$ & 2.0(2) & 0.53(4)  & 192.4(3) & 3.8(1) \\
Fe$_{0.75}$Co$_{0.25}$Sb$_2$ & 1.1(1) & 0.25(2) & 189.9(2) & 1.80(6) \\
Fe$_{0.55}$Co$_{0.45}$Sb$_2$ & 1.5(1) & 0.09(4) & 190.9(2) & 1.25(6) \\
Fe$_{0.75}$Cr$_{0.25}$Sb$_2$ & 0.8(2) & 0.43(7) & 183.2(3) & 1.93(11) \\
Fe$_{0.5}$Cr$_{0.5}$Sb$_2$ & 1.4(1) & 0.26(5) & 176.2(3) & 0.64(11) \\
\end{tabular}
\end{ruledtabular}
\end{table}

From the BWF analysis of the experimental data for Fe$_{1-x}$(Co,Cr)$_x$Sb$_2$ (solid lines at the right panel of Fig \ref{4}), one can see a large increase  of the $q$ with an increase of $x$, indicating a decrease of electron-phonon interaction by Co and Cr alloying. Temperature dependance of energy and linewidth for Fe$_{1-x}$(Co,Cr)$_x$Sb$_2$ alloys are shown in the left panel of Fig. \ref{fig4}. Experimental data are represented by symbols. Calculated spectra, obtained using Eqs. (\ref{10}) and (\ref{4}), are represented by dashed and solid lines. Best fit parameters are presented in Table \ref{tab.2}. $\Gamma_0$ decreases significantly in Fe$_{1-x}$(Co,Cr)$_x$Sb$_2$ alloys compared  to the pure FeSb$_2$ (Table \ref{tab.2}), although crystal disorder increases with increasing Co and Cr concentration (for $x\leq0.5$).  This is a consequence  of drastic decrease of electron-phonon interaction contribution with increasing $x$. One can also notice that values of anharmonic constants decrease with increasing Co and Cr concentrations, what can be a consequence of change of electronic structure of material by alloying.

\begin{figure}
\includegraphics[width=0.4\textwidth]{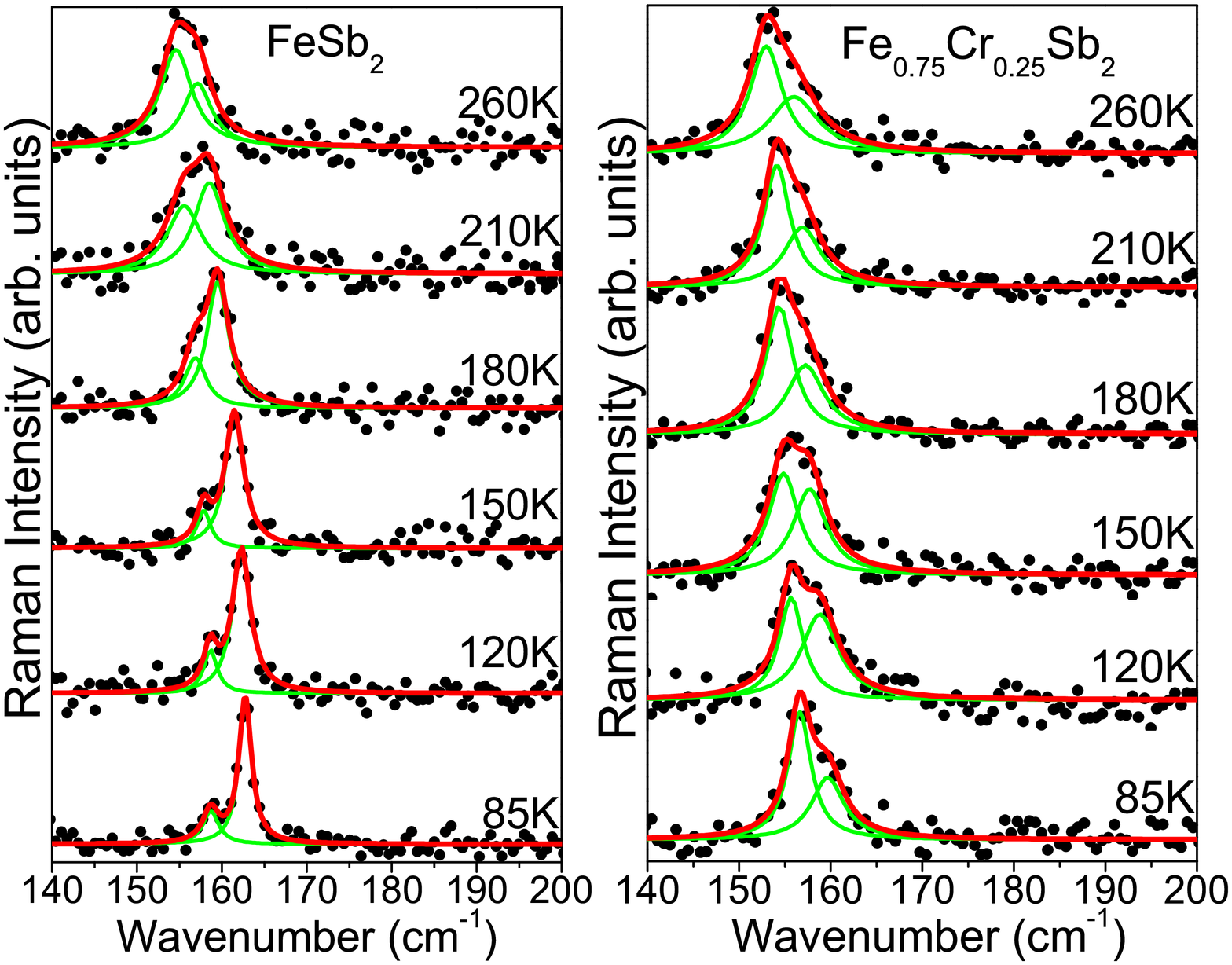}
\caption{The Raman scattering spectra of FeSb$_2$ (left
panel) and Fe$_{0.75}$Cr$_{0.25}$Sb$_2$ (right panel) single crystals in the (x$^,$x$^,$) configuration (A$_{g}$ symmetry modes) measured at different temperatures.}
\label{fig5}
\end{figure}

The polarized Raman scattering spectra for pure FeSb$_{2}$ and Fe$_{0.75}$Cr$
_{0.25}$Sb$_{2}$ single crystals in the (x$^{,}$x$^{,}$) configuration (A$_g$ symmetry modes) measured at different temperatures, are presented in Fig. \ref{fig5}. We have observed structure at about 155 cm$^{-1}$ which shows
asymmetry towards higher wavenumbers. However, this asymmetry cannot be ascribed to the electron-phonon interaction, but is a consequence of the existence of two A$_g$ symmetry modes, as we have already reported in our previously published paper.\cite{7} Low temperature measurement confirmed our previous assignation. The Lorentzian lineshape profile has been used for extraction of mode energy and linewidth. These modes have nearly the same energies what imposes the existence of the mode mixing, manifested by energy and intensity exchange. The mixing is
specially pronounced when the intensities of the modes are nearly the
same.\cite{13} For FeSb$_{2}$ the mixing is strongest in the temperature
range between 200 K and 250 K  and for Fe$_{0.75}$Cr$_{0.25}$Sb$_{2}$ between 120 K and 180 K, see Fig. \ref{fig5}.

\section{Conclusion}

The temperature study of polarized Raman scattering spectra of the Fe$_{1-x}$M$_x$
Sb$_2$ (M=Cr, Co) single crystals has been performed. The linewidths and energies of the Raman modes were analyzed as a
function of $x$ and temperature. Strong electron-phonon
interaction, observed for the B$_{1g}$ symmetry mode of pure FeSb$_2$, produces significant mode asymmetry. The coupling constant reaches highest value at about 40 K and after that remains temperature independent. Additional broadening comes from the temperature induced anharmonicity. With increasing concentration of Co and Cr in Fe$_{1-x}$(Co,Cr)$_x$Sb$_2$ alloys the electron-phonon interaction is drastically reduced. We have also observed mixing of the $A_g$ symmetry phonon modes in pure and Cr doped sample.

\section*{Acknowledgment}

We have pleasure to thank Dr Zorani Doh\v{c}evi\'c-Mitrovi\'c for helpful discussion. This work was supported by the Serbian Ministry of Science and Technological
Development under Project No. 141047. Part of this work was carried out at
the Brookhaven National Laboratory which is operated for the Office of Basic
Energy Sciences, U.S. Department of Energy by Brookhaven Science Associates
(DE-Ac02-98CH10886).

$^{\ast }$ Present address Ames Laboratory and Department of Physics and
Astronomy, Iowa State University, Ames, Iowa 50011, USA

\end{document}